\documentclass[aps,prd,twocolumn,superscriptaddress,nofootinbib,showpacs,letterpaper]{revtex4}

\usepackage{verbatim}
\usepackage[usenames,dvipsnames]{color}
\usepackage[pdftex]{graphicx}
\usepackage[english]{babel}
\usepackage{amsmath,amssymb}
\usepackage[colorlinks=true,citecolor=blue,linkcolor=magenta]{hyperref}

\begin{document}
\newcommand{\E}{\mathrm{E}}
\newcommand{\Var}{\mathrm{Var}}
\newcommand{\bra}[1]{\langle #1|}
\newcommand{\ket}[1]{|#1\rangle}
\newcommand{\braket}[2]{\langle #1|#2 \rangle}
\newcommand{\mean}[2]{\langle #1 #2 \rangle}
\newcommand{\be}{\begin{equation}}
\newcommand{\ee}{\end{equation}}
\newcommand{\ba}{\begin{eqnarray}}
\newcommand{\ea}{\end{eqnarray}}
\newcommand{\SD}[1]{{\color{magenta}#1}}
\newcommand{\rem}[1]{{\sout{#1}}}
\newcommand{\alert}[1]{\textbf{\color{red} \uwave{#1}}}
\newcommand{\Y}[1]{\textcolor{yellow}{#1}}
\newcommand{\R}[1]{\textcolor{red}{#1}}
\newcommand{\B}[1]{\textcolor{blue}{#1}}
\newcommand{\C}[1]{\textcolor{cyan}{#1}}
\newcommand{\db}{\color{darkblue}}
\newcommand{\intinfty}{\int_{-\infty}^{\infty}\!}
\newcommand{\Tr}{\mathop{\rm Tr}\nolimits}
\newcommand{\const}{\mathop{\rm const}\nolimits}
\makeatletter
\newcommand{\rmnum}[1]{\romannumeral #1}
\newcommand{\Rmnum}[1]{\expandafter\@slowromancap\romannumeral #1@}

\makeatother

\title{Effects of Mirror Aberrations on Laguerre-Gaussian 
       Beams in Interferometric Gravitational-Wave Detectors}
\author{T Hong}
\affiliation{California Institute of Technology, Pasadena, CA 91125, USA }
\author{J Miller}
\affiliation{The Australian National University, Canberra, ACT 0200, Australia}
\author{H Yamamoto}
\affiliation{California Institute of Technology, Pasadena, CA 91125, USA }
\author{Y Chen}
\author{R Adhikari}
\affiliation{California Institute of Technology, Pasadena, CA 91125, USA }

\begin{abstract}
  A fundamental limit to the sensitivity of optical interferometers is
  imposed by Brownian thermal fluctuations of the mirrors' surfaces.
  This thermal noise can be reduced by using larger beams which ``average
  out'' the random fluctuations of the surfaces. It has been
  proposed previously that wider, higher-order Laguerre-Gaussian modes
  can be used to exploit this effect. In this article, we show that
  susceptibility to spatial imperfections of the mirrors' surfaces
  limits the effectiveness of this approach in interferometers used
  for gravitational-wave detection. Possible methods of reducing this
  susceptibility are also discussed.


\end{abstract}
\maketitle

\section{Introduction}
Long-baseline laser-interferometer gravitational-wave detectors, such
as those used in LIGO~\cite{LIGO}, VIRGO~\cite{Virgo},
GEO600~\cite{GEO} and LCGT~\cite{LCGT}, use Michelson interferometry
to measure tiny differential changes in arm length induced by
gravitational waves.  Spurious motions of a mirror's surface, such as
those caused by seismic, thermal, and radiation-pressure fluctuations,
can compromise the sensitivity to gravitational wave signals.  Brownian
thermal noise in the dielectric mirror coatings, or {\it coating
  Brownian noise}, is known to be the dominant noise source in the
intermediate frequency band of Advanced
LIGO~\cite{Harry:CoatingThermal} and other similar interferometers.

As described by the Fluctuation-Dissipation
Theorem~\cite{Callen:FDT,Levin:FDT}, dissipation via internal friction in the
dielectric coatings must lead to fluctuations in the thickness of the
coatings. When the beam spot size is much larger than the coating
thickness, coating Brownian noise at different locations on the
mirror's surface can be considered to be
uncorrelated. This leads to the following
scaling law~\cite{Lovelace:thermal, O'Shaugnessy:beam},
\begin{equation}
\label{eq:ScalingLaw}
S_x \propto \frac{\displaystyle \int  I^2(\vec r) \,\mathrm{d}^2 \vec r}
                          {\displaystyle \left[\int I(\vec r)  \,\mathrm{d}^2 \vec r \right]^2}
\end{equation}
which describes how the power spectrum of observed coating Brownian
noise $S_x$ depends on the intensity profile $I(\vec r)$ of the
optical field which is used to read out the mirror motion;
i.e.~the coating Brownian noise power spectrum is inversely 
proportional to the effective area of the optical mode.
\begin{table}[htbp!]
  \caption{Beam shapes that have been considered for use in gravitational-wave detectors, mirror shapes that support them and their thermal-noise suppression factors (in power) for Advanced LIGO parameters (cavity length $L=4$~km, mirror radius of 17~cm).}
\begin{ruledtabular}
\begin{tabular}{cccc}
Mode        & Mirror Shape & Suppression Factor & Ref.\\
\hline
LG$_{3,3}$   & Spherical    & 1.61               & \cite{LG}\\
Mesa        & Sombrero     & 1.53               & \cite{Bondarescu:Mesa,Miller:Thesis}\\
Conical     & Conical      & 2.30               & \cite{conical}
\end{tabular}
\end{ruledtabular}  
\label{tab:modes}
\end{table}

Three families of optical modes have so far been considered for
mitigating coating thermal noise (see Table~\ref{tab:modes}).  Among
these modes, only the higher-order Laguerre-Gauss mode, LG$_{3,3}$,
can be supported by optical cavities employing standard spherical
mirrors. Due to the practical advantages associated with the use of
spherical mirrors, experimental testing of LG$_{3,3}$ modes has
begun. It has thus far been demonstrated that these modes can be
generated with high efficiency and resonated in tabletop cavities with
small mirrors~\cite{Fulda:LG,Matteo:LG}.

An unpleasant property of higher-order LG modes is that each
LG$_{p,l}$ mode is $2p+|l|+1$-fold degenerate, the LG$_{3,3}$ mode
being 10-fold degenerate. Mirror figure errors will inevitably split
each formerly degenerate mode into $2p+|l|+1$ single modes with
eigenfrequencies which depend on the particulars of the figure
error. By contrast, a non-degenerate mode, under the same figure
error, will usually remain as a single, weakly-perturbed
non-degenerate mode.

In this work we explore the effects of LG$_{3,3}$ modal degeneracy
quantitatively, via both numerical and analytical methods. Guided by
experience with existing interferometers we have selected contrast
defect as our metric of interferometer performance.

To ground our investigation in reality we incorporate mirror figure
errors derived from measurements of the first Advanced LIGO
optics. The creation of these maps is described in
Section~\ref{sec:Mirrorphasemap}.

In Section~\ref{sec:nocorrection}, we use perturbation theory to
analyze the effect of such mirror perturbations on the degenerate
subspace which includes the LG$_{3,3}$ mode.  We then use this newly
perturbed set of modes to calculate the contrast degradation of a
single Fabry-Perot arm cavity analytically in
Section~\ref{sec:AnalyticalContrastDefect}.

In Section~\ref{sec:Numerical}, we utilize a sophisticated numerical
field propagation code to confirm analytical results and examine a
more complicated interferometer topology.

In Sec.~\ref{sec:Improvement}, we explore two methods of mitigating
contrast degradation; neither of the methods was ultimately
successful.

\section{ Mirror Figure Errors}
\label{sec:Mirrorphasemap}
\begin{figure}[htbp!]
\begin{center}
\includegraphics[width=\columnwidth]{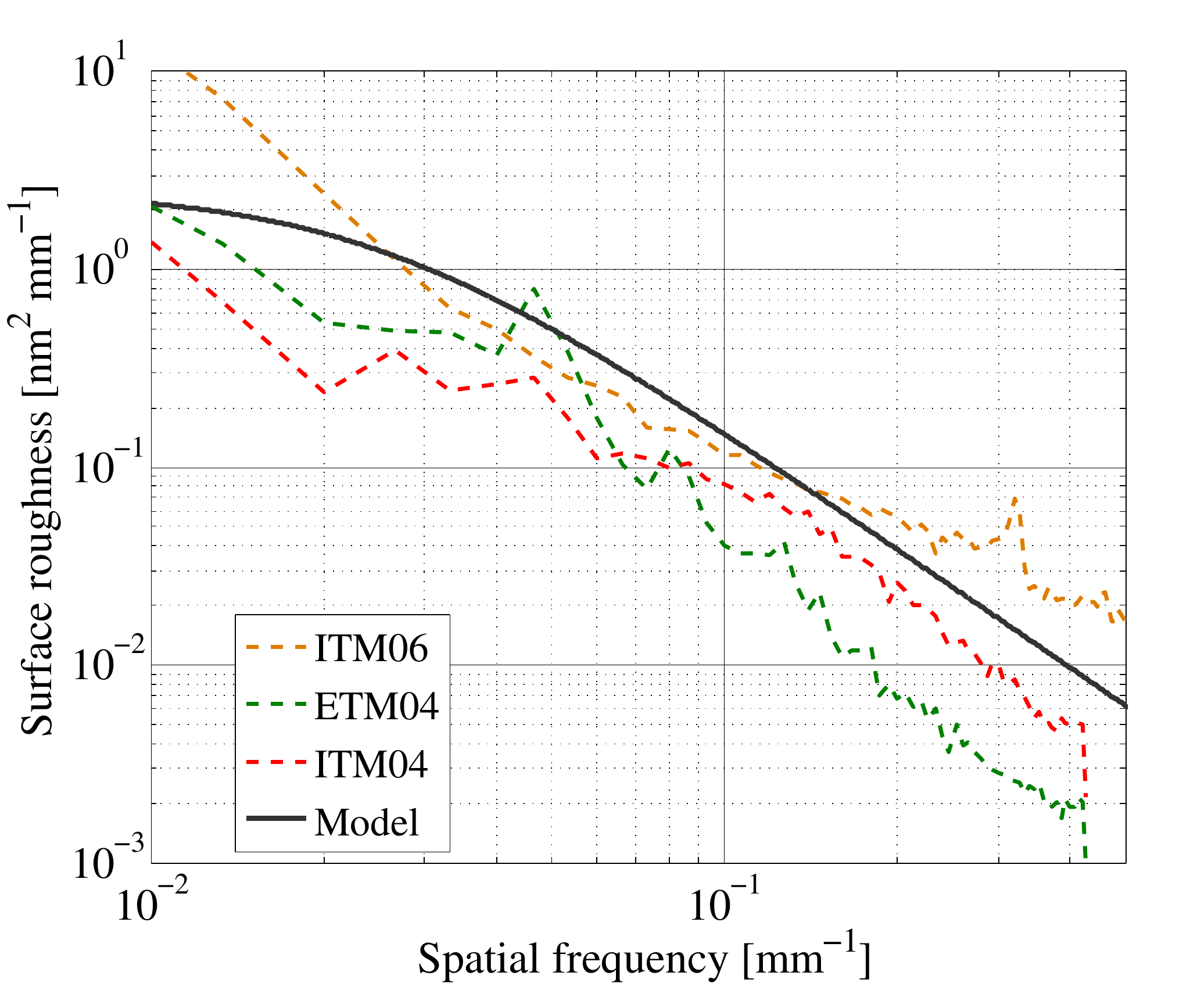}
\caption{Power spectral densities of uncoated mirror surface
  roughness.  The dashed lines are the measured spectra of three
  Advanced LIGO arm cavity mirrors. A model approximating these
  spectra (black trace, see~(\ref{eq:spectraModel})) was created to
  generate the random maps used in our work.}
\label{fig:surfacePSDs}
\end{center}
\end{figure}
In this work we investigate the consequences of realistic mirror
imperfections on the performance of the LG$_{3,3}$ mode. The
parameters of the particular imperfections applied are therefore
significant.


Fig.~\ref{fig:surfacePSDs} illustrates measured surface roughness
power spectral densities of selected Advanced LIGO mirror substrates,
prior to the application of dielectric mirror coatings (the PSD plots end up to spatial 
wavelength of 2 mm). 
Based on the measured Initial LIGO optics and small optics of Advanced LIGO, we construct an analytical model (solid black line) which falls roughly in the middle of Advanced LIGO test mass PSDs,
\begin{equation}
\label{eq:spectraModel}
   S(f)\propto(1+(0.04f)^2)^{-1}.
\end{equation}
This one-dimensional function was used to generate random mirror maps
which are statistically similar to those one might find in an advanced
gravitational-wave interferometer. Such random mirror maps, used in
all aspects of this investigation, were constructed by multiplying
each point of the amplitude spectral density's magnitude by a random
complex number $a+ib$ before transforming back to coordinate space and
appropriately scaling the result to yield the desired RMS. Scalars $a$
and $b$ are drawn independently from a normal distribution with zero
mean and a standard deviation of one \cite{Bondu}. 

The entire surface is fit by Zernike polynomials and the terms corresponding to Piston, tilt and
power (Zernike polynomials $Z_0^0$, $Z_1^{\pm1}$ $Z_2^0$) were
removed from our maps (the Piston term is irrelevant because the lock process adjusts the microscopic length; the tilt term is removed to represent the
alignment control; the ROC of the generated surface is corrected by hand). The RMS values quoted are calculated after this
subtraction.

\begin{figure}[thbp!]
  \includegraphics[width=\columnwidth]{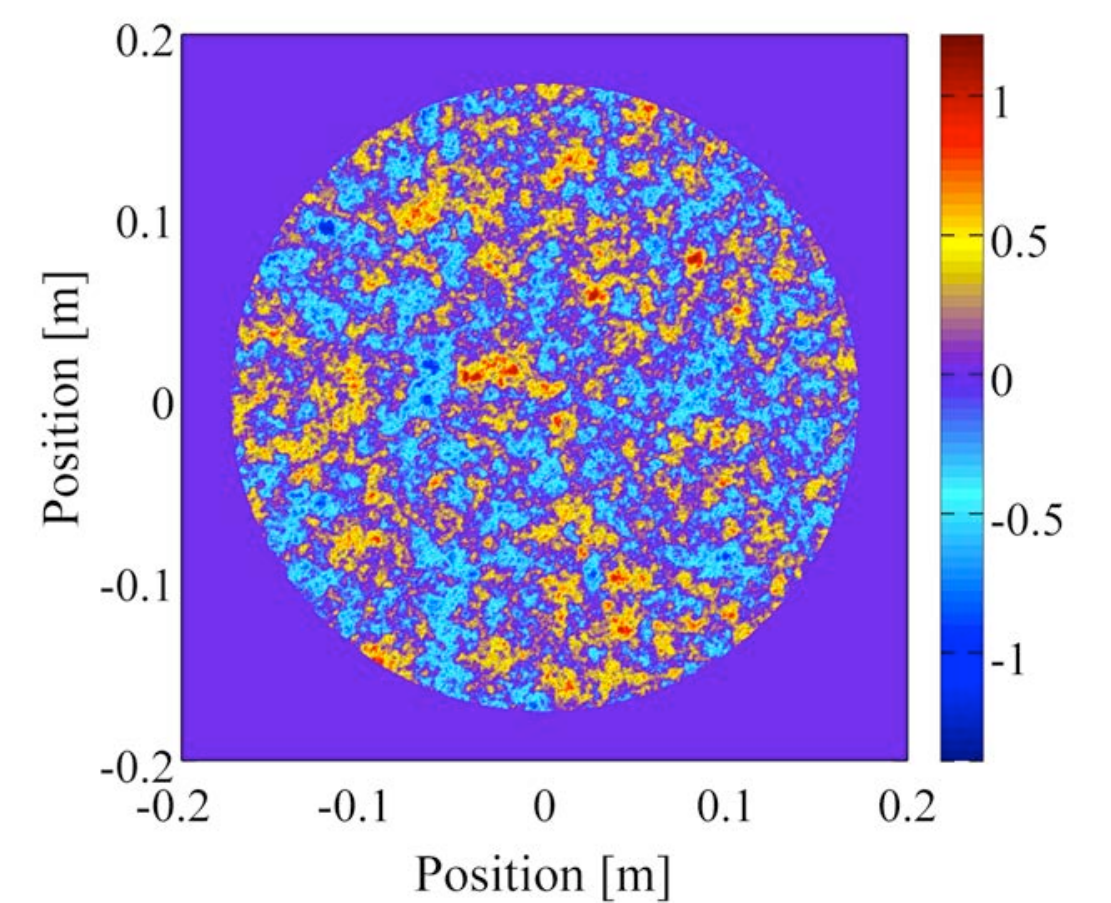}
  \caption{Surface figure (in nm) of a typical generated phase map with Piston, tilt, power and
    astigmatism terms subtracted.}
 \label{fig:MirrorPhaseMap}        
\end{figure}

Fig.~\ref{fig:MirrorPhaseMap} shows the surface figure of one map
generated using our algorithm. This map is typical of a larger
population and was selected as a reference to be used in all
analytical calculations.
\section{Degenerate Perturbation-Theory Analysis}
\label{sec:nocorrection}

\subsection{Laugerre-Gauss modes}
The Laguerre-Gauss modes (LG$_{p,l}$) are a set of circularly
symmetric modes which can be written in cylindrical coordinates
as~\cite{Siegman}
\begin{equation}
  \begin{split} 
    u_{p,l}(r,\phi,z)&=\sqrt{\frac{2\,p!}{\pi
        (|l|+p)!}}\frac{1}{\omega(z)}\left[\frac
      {\sqrt 2 r}{\omega(z)}\right]^{|l|}\\
    &\times L_p^{|l|}\left[\frac{2r^2}{\omega^2(z)}\right]  \exp[i(2p+|l|+1)\psi(z)]\\
    &\times \exp\left[-i k\frac{r^2}{2R(z)}+il\phi\right]\exp\left[\frac{-r^2}{\omega^2(z)}\right],
    \end{split}
\end{equation}
where $\omega(z)$ is the beam radius, $\psi(z)$ is the Gouy phase, and
$R(z)$ is phase front curvature of the beam. 
$L_p^{|l|}(x)$ is the associated Laguerre polynomial where $p\geq0$ and
$l$ are the radial and azimuthal indices respectively.

The mode selectivity of the cavity is determined by the cavity finesse
and the mode dependent phase shift $(2p+|l|+1)\psi(z)$. From this we
see that the LG$_{p,l}$ mode has $2p+|l|+1$ degenerate eigenmodes.
For example, LG$_{3,3}$ belongs to a 10-fold degenerate space, which can be spanned by: LG$_{3,\pm3}$,
LG$_{0,\pm9}$, LG$_{1,\pm7}$, LG$_{2,\pm5}$ and LG$_{4,\pm1}$.
\begin{figure}
\includegraphics[width=\columnwidth]{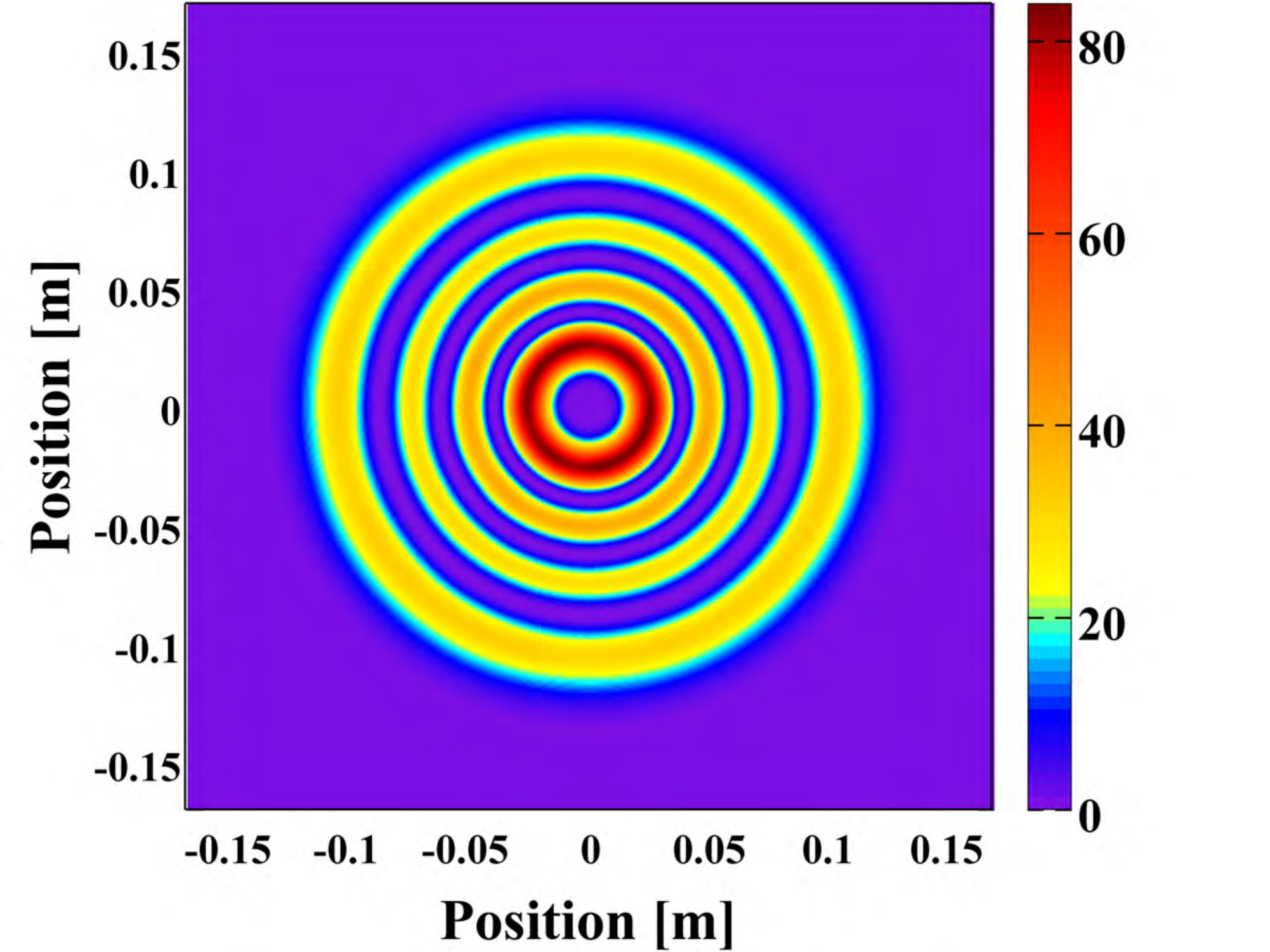}
\caption{Normalized intensity distribution of the $\rm{LG}_{3,3}$ mode
  at the mirror position ($\omega_0=0.021$ m, $z=1997.25$~m).}
\label{fig:LG33}
\end{figure}

Coating Brownian noise power is proportional to the integral of beam
intensity, as the scaling law (Eq.~(\ref{eq:ScalingLaw}))
indicates. In Table~\ref{table:BeamClipping}, we present theoretical
thermal noise suppression factors for selected LG modes. To permit a
fair comparison, the widths of all modes considered here and
henceforth were chosen to be $0.018$~m, which yield a clipping
loss~\cite{Freise:LG}, due to the finite size of the cavity mirrors,
of around 1~ppm. We see that the LG$_{3,3}$, considered by many as the
leading candidate for use in gravitational wave interferometers,
offers a theoretical thermal noise reduction factor of $\sim$1.6
compared to a standard Gaussian beam (LG$_{0,0}$). The transverse
intensity distribution of the LG$_{3,3}$ mode is presented in
Fig.~\ref{fig:LG33}.
\begin{table}
  \caption{Suppression factors of thermal noise (in power spectral density) for LG modes with a fixed clipping loss of 1
    ppm. }
\begin{tabular}{lccccccccccccc}
\toprule
   & & ${\rm LG_{0,0}}$ & &${\rm LG_{0,9}}$ && ${\rm LG_{1,7}}$ & &${\rm LG_{2,5}}$& &${\rm LG_{3,3}}$& &${\rm LG_{4,1}}$ \\
\hline
 Beam radius\\(mm) && $9.96$ && 16.5&& 17.3 && 17.9&&18.2&& 18.4\\
\hline
Suppression\\
Factor &&1 &&1.51&&1.62&&1.64&&1.61&&1.51\\
\toprule
\end{tabular}
\label{table:BeamClipping}
\end{table}

\subsection{Application of Degenerate Perturbation Theory to
                     the Perturbed Fabry-Perot Cavity}
\label{sec:PerturbTheory}
The combination of eigenmodes excited in a cavity depends on the
composition of the incident field and on the properties of the cavity
itself. In this section we discuss how first-order perturbation theory
can be applied to this problem. We first explore how mirror figure
error breaks the degeneracy of LG cavity modes before describing the
phase shift each mode experiences in an optical cavity and finally
constructing the total field (prompt plus leakage) reflected from a
perturbed Fabry-Perot resonator.

\subsubsection{Mode Splitting}
\begin{figure}[htbp!]
  \centering
   \includegraphics[width=\columnwidth]{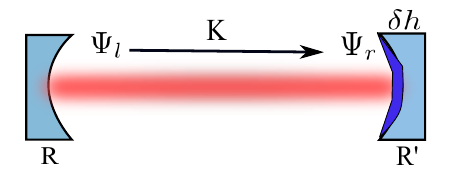}
   \caption{Fabry-Perot cavity with a perturbation $\delta h(x,y)$ on the
     end mirror.}
  \label{fig:FPcavity}
\end{figure}

Fig.~\ref{fig:FPcavity} illustrates light propagation in a simple
Fabry-Perot cavity, introducing the notation employed in our
formalism.

We use the following standard method for propagating cavity
fields,
\be
\Psi_{r}(\vec r)=\int K(\vec r,\vec{r}\,^\prime)\Psi_l(\vec{r}\,^\prime)\,\mathrm{d}^2\vec{r}\,^\prime,
\ee
or $\ket {\Psi_r}=\hat K \ket {\Psi_l}$, where $\Psi_{r}$ and
$\Psi_{l}$ are the electric fields near the right and left mirror
respectively and $K$ is the field propagator from the left mirror to the
right
\ba
K(\vec r,\vec{r}\,^\prime)=\frac{i k}{2\pi L}e^{-\frac{i k}{2L}\vert \vec r-\vec{r}\,^\prime\vert^2}.
\ea

Therefore, if $\ket \psi$ is an eigenmode of the cavity,
\be
\ket \psi=\hat R\hat K\hat{R}^\prime\hat K\ket \psi,
\ee
where $\hat R$ is the reflection operator of the left mirror and
$\hat{R}^\prime$ is for the right. Hence, $\ket \psi$ is an eigenmode of
the operator $(\hat R\hat K\hat {R}^\prime\hat K)$.

We assume that the mirror on the left is ideal and study the
consequences of applying a surface figure perturbation $\delta h$ to
the right hand (end mirror) optic. The reflection operator $\hat
R^\prime$ can then be written as \be
\label{RR}
\hat {R'}=\hat{R}e^{2ik\delta h}\approx \hat{R}(1+2ik\delta h).
\ee
To obtain the real cavity eigenmodes, we need to solve for the 
eigenfunctions of $(\hat R\hat K\hat{R}^\prime\hat K)$.
Note that the original LG$_{33}$ mode is an eigenfunction of 
an unperturbed cavity:
\be
\ket{33}=(\hat R\hat K\hat {R}\hat K)\ket{33}.
\ee

As introduced above, the LG$_{33}$ mode is degenerate with $9$ other
modes, each having eigenfrequency $\omega_0$, thus $\hat R\hat K\hat
{R}\hat K$ also has $10$ degenerate eigenmodes at $\omega_0$.

For reasonable parameters, modes outside of this degenerate sub-space
are far enough from resonance that we may ignore them in this
first-order analysis. Therefore, to good approximation, we can assume
the new eigenmodes of the perturbed cavity are still members of the
Hilbert space of the original $10$ LG modes, which we represent by
$\ket i$, (i=1,2,\ldots,10). These new eigenmodes, denoted
$\ket{i^\prime}$, are the eigenvectors of the matrix with elements
$\bra{i}\hat R\hat K\delta h\hat R\hat K\ket{j}=\bra i\delta h\ket j$,
$(i,j=1,2,\ldots,10)$ \ba
\label{eq:matrix}
\begin{pmatrix}
\bra 1\delta h\ket 1 & \ldots & \bra 1\delta h\ket {10} \\
\vdots & \ddots & \vdots \\
\bra {10}\delta h\ket 1 & \ldots & \bra {10}\delta h\ket {10}
\end{pmatrix}.
\ea

Denoting the electric field of $\ket i$ as $\phi_i$, we write
\be
\bra i\delta h\ket j=\int\int \phi^*_i(x,y)\delta h(x,y)\phi_j(x,y)\, \mathrm{d}x\,\mathrm{d}y.
\ee
The frequency shift of the degenerate modes introduced by the
perturbation is then proportional to the eigenvalues of the matrix
\be
\label{omegai}
\omega_i =\frac{k c}{L} \bra{i'}\delta h\ket{i'},
\ee
where $k=2\pi/\lambda$ is the optical wavenumber, $c$ is the speed of
light and $L$ is the cavity length.

\begin{figure}[htbp!]
   \includegraphics[width=\columnwidth]{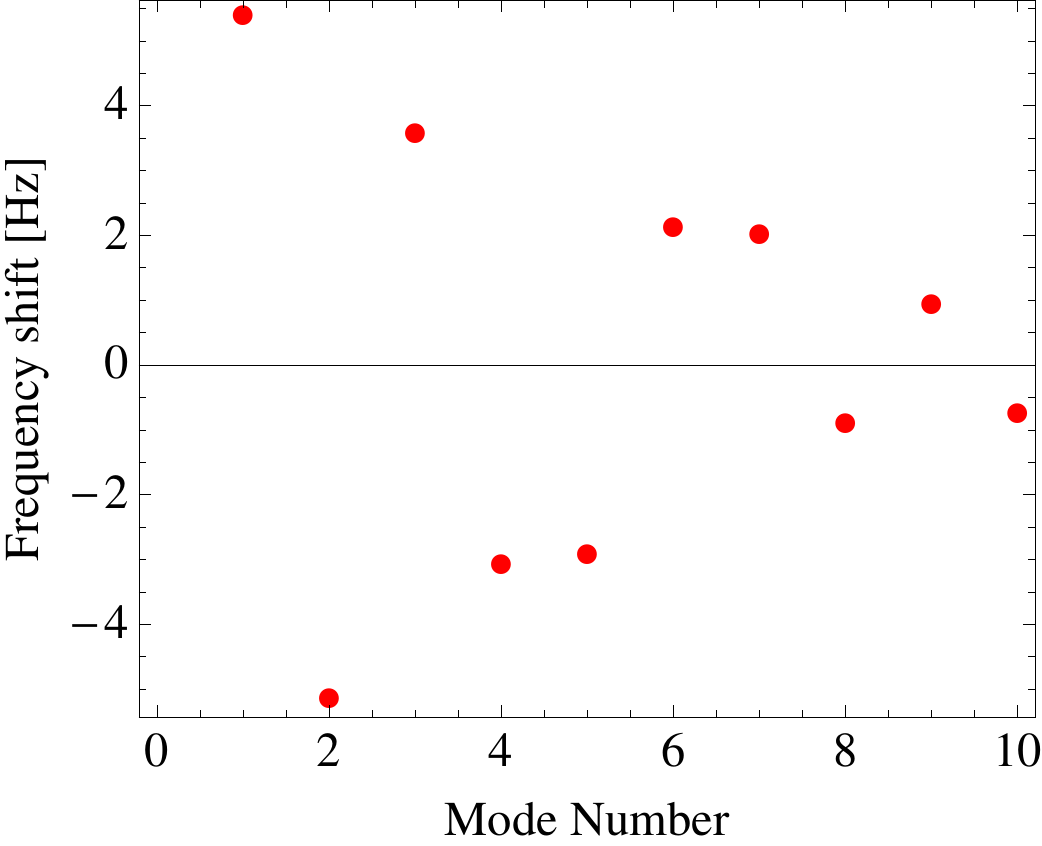}
     \caption{Frequency shift of LG modes introduced as a result of
                   realistic mirror perturbations.}
   \label{fig:split}
\end{figure}
This quantity was evaluated using the reference map shown above
(Fig.~\ref{fig:MirrorPhaseMap}). The RMS roughness of the reference
was scaled to be similar to those of measured Advanced LIGO mirror
surfaces (0.3~nm~RMS). Results are presented in
Fig.~\ref{fig:split}. Here the frequency splits are given in Hertz
($\omega/2\pi$). The frequency shifts are one order of magnitude
smaller than the advanced gravitational-wave interferometer's cavity
linewidth. Thus, multiple perturbed eigenmodes will be partially
resonant, radically distorting the shape of the output field.

\subsubsection{The Modal Input-Output Equation}
Above we have shown that mirror figure errors will lift the modal
degeneracy and split the degenerate LG$_{3,3}$ space into distinct
states with unique eigenfrequencies. We now consider how each of these
modes interacts with a cavity.

For any mode $\ket{\rm in}$ injected into an ideal cavity, there
exists a frequency dependent phase shift between the input and total
reflected or {\it output} fields.  This can be written
as~\cite{Milburn} \be
\label{eq:inoutrelation}
\ket{{\rm out}}=\frac{\gamma_c+ i (\omega - \omega_0)}{\gamma_c-i
  (\omega-\omega_0)}\ket{{\rm in}}, \ee where $\omega$ is the
frequency of the injected field, $\omega_0$ is the resonant frequency
of the cavity closest to $\omega$ and
\hbox{$\gamma_c=c{\rm T_\mathrm{input}}/4L$} is the cavity pole frequency in
Hertz. Here, ${\rm T_\mathrm{input}}$ denotes the power transmissivity of the
cavity input mirror; the transmissivity of the end mirror is assumed
to be zero.

Suppose that this ideal cavity hosts an $N$-fold degenerate space and
that we inject an input mode $|\rm in\rangle$ which belongs to this
space. If the $N$-fold degeneracy is broken by some mirror figure
error, the new eigenmodes can be approximated by $|n\rangle$,
$n=1,2,\ldots, N$, where each mode still belongs to the original
sub-space, but has a new eigenfrequency $\omega_n$.  As we shall see
in Appendix~\ref{sec:Equiv}, this is justified as long as the cavity
finesse is high enough and the eigenfrequencies of the non-degenerate
modes are well-separated from this sub-space.

The output from the perturbed cavity can be obtained by projecting the
input mode $|\rm{in}\rangle$ onto the new basis $|n\rangle$ and
calculating the phase shifts using the following relation:
\be
\ket{{\rm out}}=\sum_{n'=1}^N \frac{\gamma_c+i (\omega-\omega_{n'})}{\gamma_c-i (\omega-\omega_{n'})}\ket{n'} \braket{n'}{{\rm in}}.
   \label{eq:output}
\ee
\begin{figure}
   \includegraphics[width=\columnwidth]{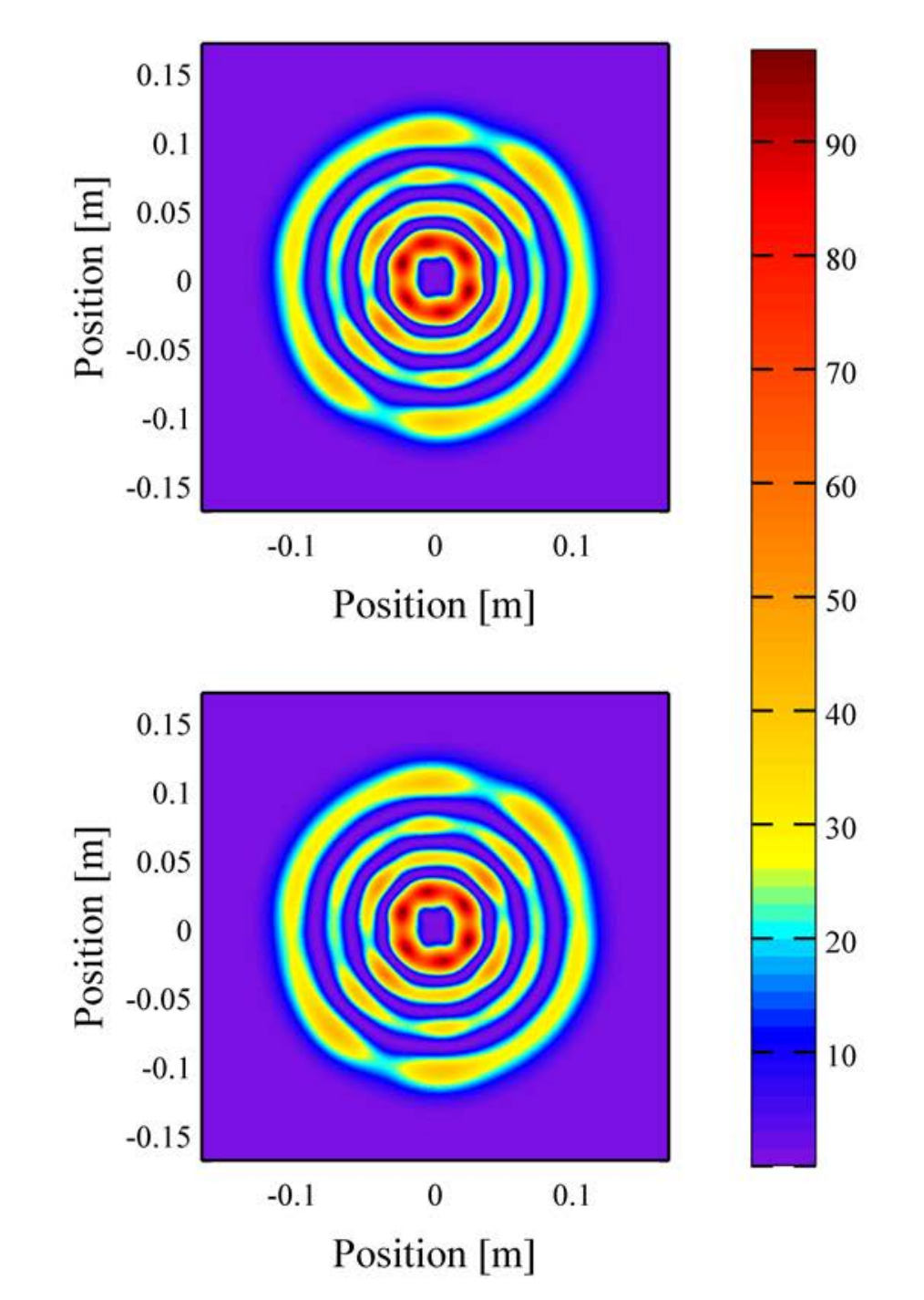}
   \caption{Intensity distributions of the total field reflected from
     an arm cavity whose end mirror was perturbed by the reference
     map. These distributions were calculated independently via two
     different techniques. Top -- Analytic method described in
     Section~\ref{sec:PerturbTheory}. Bottom -- FFT-based numerical
     simulation (see Section~\ref{sec:Numerical}).}
    \label{fig:output}
\end{figure}

This procedure was applied to the LG$_{3,3}$ degenerate space using
the resonant frequencies of the previous section, $\omega_n$. With
the reference phase map scaled to 0.3~nm~RMS, the intensity profile of
the resulting output mode is shown in Fig.~\ref{fig:output}.

\section{Contrast Defect}
\label{sec:Contrast}
In gravitational-wave interferometers, contrast defect is defined as
the ratio of the minimum possible optical power at the anti-symmetric
(dark) port to the power incident on the beamsplitter. This quantity
can be expressed as
\begin{equation}
C=\frac{P_\mathrm{AS}}{P_\mathrm{X}+P_\mathrm{Y}},
\label{eq:NumCD}
\end{equation}
where X and Y are labels for the two arm cavities, $P_\Box=\iint_{\mathbb{R}^2}(\Psi^\mathrm{out}_\Box)^*\Psi^\mathrm{out}_\Box\,\mathrm{d}x\mathrm{d}y$
and
$\Psi^\mathrm{out}_\mathrm{AS}=\Psi^\mathrm{out}_\mathrm{X}-\Psi^\mathrm{out}_\mathrm{Y}$
represents the field at the anti-symmetric port (AS) of the interferometer.

In principle, the dark port could be completely dark. However, the
presence of intentional imbalances in the arms (finite beamsplitter
size, Schnupp asymmetry, etc.)  and unintentional imperfections
(mirror shape, scatter loss, mirror motion, etc.) result in imperfect
destructive interference between the fields which recombine at the
beamsplitter. This imperfect interference leads to the leakage of some
`junk' light to the dark port where the gravitational-wave signal is
also detected. Excess light at the dark port can lead to a degradation
of sensitivity via several mechanisms and compromise the robust
operation of interferometer longitudinal and alignment control systems
\cite{Sigg:ReadoutControl, Nic:Contrast}. Contrast defect is thus a
useful metric to employ when comparing interferometer configurations.

The above perturbation analysis shows that the fields resonating in
the arm cavities of a real interferometer will no longer be pure
LG$_{3,3}$ modes. Further, the relative amplitudes of the
quasi-degenerate modes are strongly dependent on mirror properties,
which will, in general, be different for each arm. Hence the perturbed
arm cavity fields will interfere imperfectly at the beamsplitter. We
therefore expect an LG$_{3,3}$ interferometer to exhibit a larger
contrast defect than e.g.~an LG$_{0,0}$ mode. We now test this
hypothesis by analytical and numerical means.

\subsection{Analytic calculation}
\label{sec:AnalyticalContrastDefect}
According to Appendix~\ref{sec:Equiv}, the contrast defect in an
interferometer with one perfect arm cavity and one perturbed arm
cavity, can be analytically written as \be
\epsilon=1-|\braket{\rm{in}}{\rm{out}}|,
\label{eq:ContrastDefect}
\ee when the mirror perturbations are sufficiently small
\hbox{($\epsilon\ll1$)}.

Appendix~\ref{sec:Equiv} also shows that when a frequency shift is
small compared with the line width of the cavity,
i.e.\ $\omega-\omega_n\ll \gamma$, Eq.~\ref{eq:ContrastDefect} can be
approximated as 
\be
\epsilon = \left[\sum_{n'} \bra{33}\delta h\ket{n'}\bra{n'}\delta h\ket{33}-\bra{33}\delta h\ket{33}^2 \right]\left(\frac{8\sqrt 2 \pi}{\lambda {\rm T}}\right)^2,
\label{eq:ContrastDefectApprox}
\ee
where
\begin{equation}
\langle 33 |\delta h| pl\rangle =\int   \delta h (u_{33}^* u_{pl})\, \mathrm{d}x\,\mathrm{d}y.
\end{equation}
Alternatively, this can be written as
\begin{equation}
\epsilon = \left[\frac{(8\sqrt 2 \pi)}{\lambda T}\right]^2 \delta z^2,
\end{equation}
where
\begin{equation}
\delta z^2 = \sum_{2p+|l|=10\atop (p,l)\neq (3,3)} |\langle 33 |\delta h|pl\rangle|^2\,.
\end{equation}

This suggests that in order to minimize contrast defect, one must
strive to suppress the projection of $\delta h$ onto the 9 complex
basis functions, given by $u_{33}^* u_{pl}$.  Since these functions
are complex and $\delta h$ is real, this actually corresponds to 18
basis functions.

\begin{figure}
   \includegraphics[width=\columnwidth]{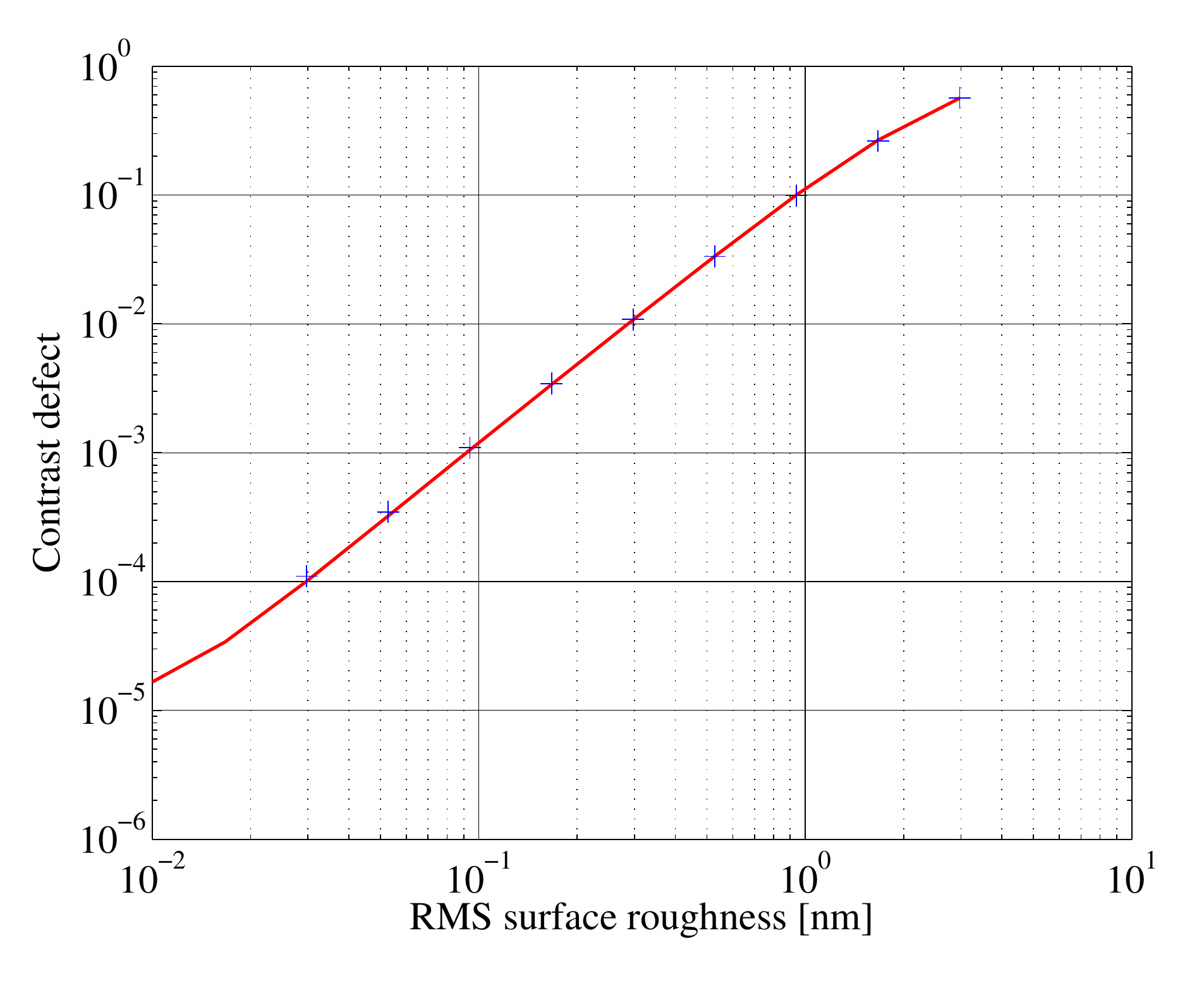}
   \caption{Contrast defect with the ETM of one cavity
     perturbed by rescaling the reference phase map: Solid red is from the
     analytical calculation, blue marker is from the FFT calculation.}
    \label{fig:comparisonplot}
\end{figure}

Based on Eq.~(\ref{eq:output}), the contrast defect has been evaluated
analytically in the case that only the end mirror (ETM) of one cavity
is perturbed with the reference phase map of
Fig~.\ref{fig:MirrorPhaseMap}.  The perturbation for different RMS
values was obtained by rescaling the reference phase map. The results
are shown in Fig.~\ref{fig:comparisonplot}.

One can show that, for small perturbations, perturbing two cavity
mirrors using phase maps derived from the same power spectral density
function, will, on average, result in twice the contrast defect when
compared to perturbations of a single mirror. However, depending on
the spatial correlations between the mirrors, the contrast defect can
be as much as twice the average in some cases.

\subsection{Numerical calculation}
  \label{sec:Numerical}

To confirm the results obtained via perturbation theory and to extend
our analysis to more complicated configurations, a parallel
investigation was carried out using numerical methods. We utilized an
FFT-based field propagation tool -- the Stationary Interferometer
Simulation (SIS)~\cite{SIS}.

SIS is predominantly used to inform the design of the Advanced LIGO
interferometers and is under continuous development at Caltech's LIGO
Laboratory. SIS employs an iterative procedure to find the stationary
fields for a given optical configuration and input beam. Mirror
surface maps can be generated from user-defined power spectral density
functions, allowing one to study the effects of various hypothetical
mirror aberrations. Cavity systems are `locked' using a
Pound-Drever-Hall signal~\cite{Drever} to realize an operating
condition similar to that which would be observed experimentally.

SIS was used to model Advanced LIGO Fabry-Perot arm cavities
supporting LG$_{0,0}$, LG$_{3,3}$ and nearly-concentric mesa
modes~\cite{Bondarescu:Mesa,Miller:Thesis}. The parameters of the
LG$_{3,3}$ and mesa cavities (see Table~\ref{table:SISparams}) were
adjusted to yield systems with round-trip diffraction loss equivalent
to that of the fiducial LG$_{0,0}$ resonator ($\lesssim1$~ppm). In
each case the input beam remained fixed as the beam which was ideally
coupled to an unperturbed cavity. To suppress the aliasing effect, a larger FFT grid of 1024$\times$1024
points on a 0.7~m$\times$0.7~m square was used for all modes.
\begin{table}[htbp!]
   \caption{Cavity parameters used in the numerical simulations. All three
    resonators had a length of 3994.5~m. The mesa radii of curvature
    refer to the fiducial sphere from which the mesa correction
    profile is subtracted (c.f.~\cite{Bondarescu:Mesa}). Cavity g
    factor is not well defined for mesa modes.}
   \label{table:SISparams}
  \begin{ruledtabular}
    \begin{tabular}{llll}
&LG$_{0,0}$&LG$_{3,3}$&Mesa\\
\hline
R$_{\mathrm{itm}}$&1934 m&2857 m&1997.25 m\\
R$_{\mathrm{etm}}$&2245 m&2857 m&1997.25 m\\
Cavity g factor&0.83  &0.16  &\hspace{1em}---
      \end{tabular}
   \end{ruledtabular}
\end{table}

\begin{figure}
     \includegraphics[width=\columnwidth]{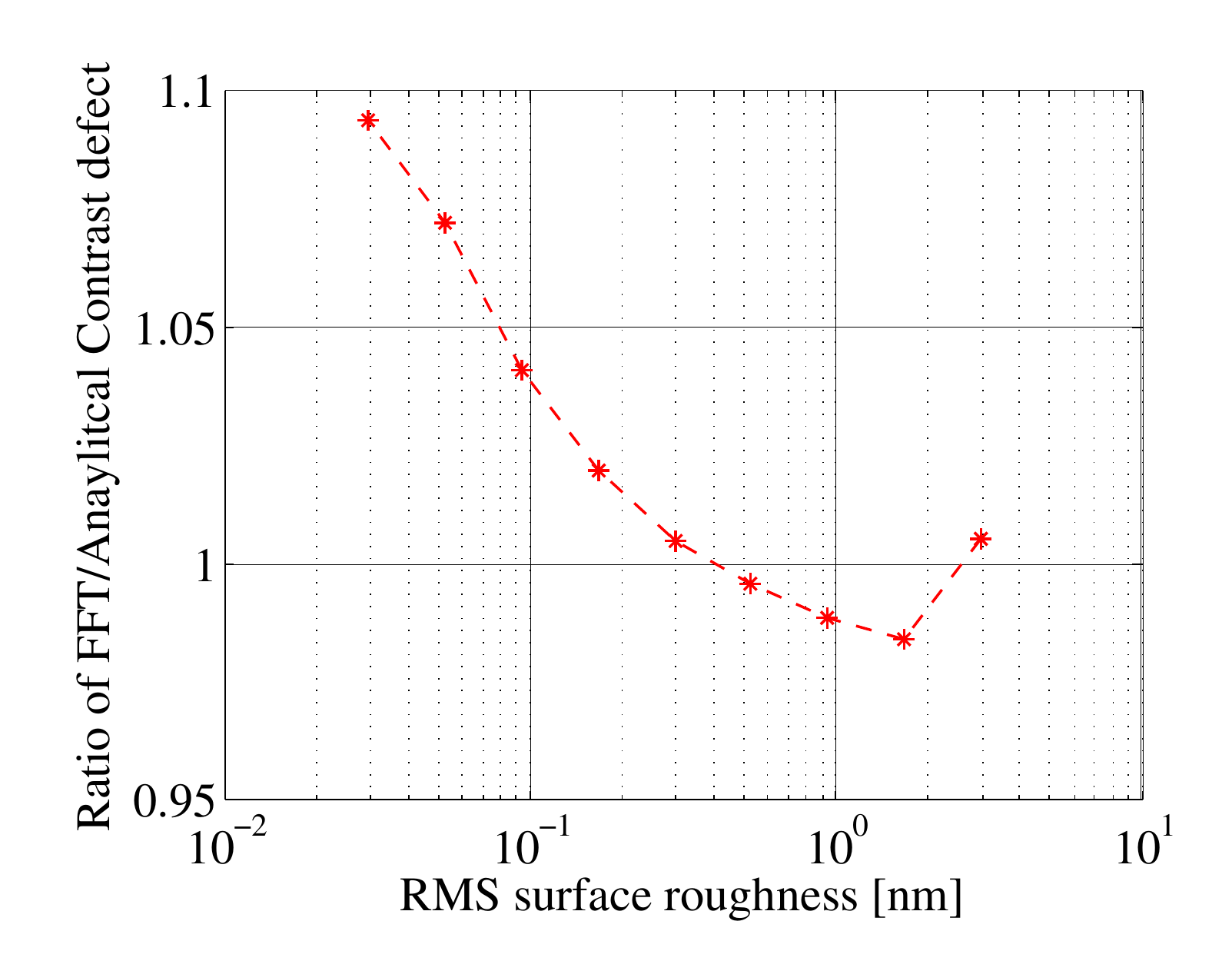}
   \caption{Ratio, numerical/analytical, of single-cavity contrast
      defects calculated with the end mirror perturbed by a 0.3~nm RMS
      figure error.}
    \label{fig:ratioplot}
\end{figure}
Initially SIS was used to simulate a configuration identical to that
studied analytically. Under these conditions both sets of results are
in good agreement (better than 10\%) (see
Figs.~\ref{fig:output},~\ref{fig:comparisonplot} and
\ref{fig:ratioplot}). The flexibility of SIS was then used to consider
more complex simulations where both cavity mirrors are perturbed and
to emulate a Fabry-Perot Michelson interferometer. SIS was also
employed to study high-RMS cases in which the analytic approximation
breaks down.

For each simulation run, two random surfaces, with a specified RMS
roughness and a spatial spectrum approximating that of the first
Advanced LIGO mirrors (see Fig.~\ref{fig:surfacePSDs}), were generated
and added to the profiles of the cavity mirrors. SIS then evaluated
the field $\Psi^\mathrm{out}$ reflected from the cavity at its
operating point, which is chosen from the cavity which is locked by using the Pound-Drever-Hall error signal.

\begin{figure}[htbp!]
  \begin{center}
  \includegraphics[width=\columnwidth]{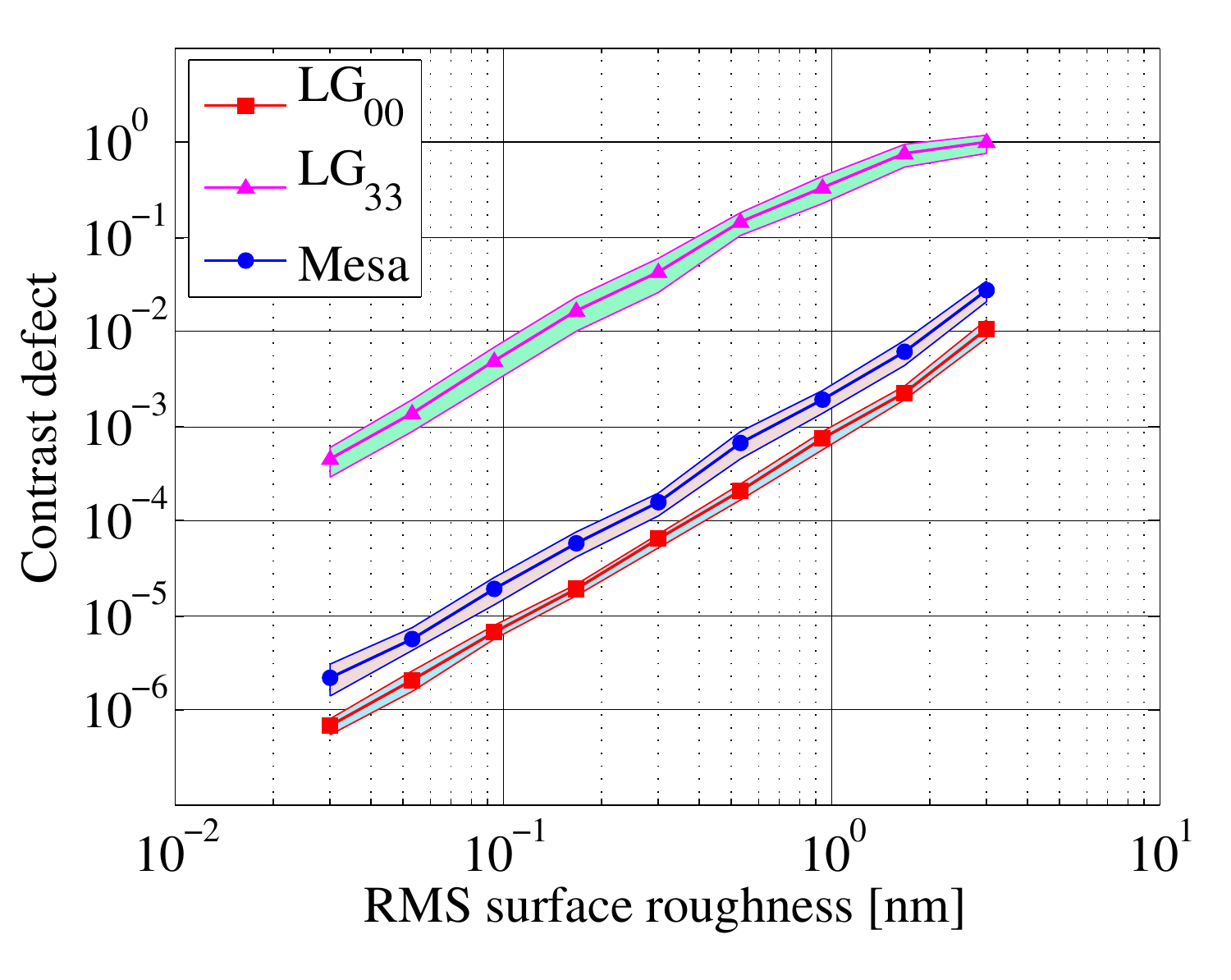}
  \caption{Interferometer contrast defect as a function of test mass
    surface roughness (all 4 mirrors are perturbed with random phase map at the same level of RMS). Solid markers report mean values of numerical
    results with the corresponding shaded regions illustrating one
    standard deviation (see~\ref{sec:Numerical}), which is roughly four times higher than the trace in Fig.~\ref{fig:comparisonplot}}
   \label{fig:ContrastDefect}
  \end{center}
\end{figure}

Results from two discrete, single-cavity simulations, representing the
$X$ and $Y$ arms of an interferometer, were then combined according to
Eq.~\ref{eq:NumCD} to estimate interferometer contrast
defect. Multiple trials were conducted at each value of RMS surface
roughness with different random maps, allowing one to consider more
than 100 unique arm cavity pairs. From these data, the mean and
standard deviation of interferometer contrast defect were found as a
function of mirror aberration RMS. Results for all three beams are
shown in Fig.~\ref{fig:ContrastDefect}. The simulated contrast defect for 
Gaussian beam ({\rm TEM00}) is consistent with the measured value of 
LIGO, which is around $10^{-4}$ ~\cite{LIGOcon} and low enough for the effective detection.

Our numerical work confirms the result from perturbation theory;
LG$_{3,3}$ interferometers are more sensitive to mirror surface roughness
than those supporting a fundamental Gaussian mode.  We further show
that LG$_{3,3}$ beams are also outperformed in this respect by
nearly-concentric mesa beams, indicating that this sensitivity arises
due to the properties of the Laguerre-Gauss mode itself and is not an
inevitable handicap for all beams capable of mitigating mirror thermal noise.

\section{Contrast Defect Improvement}
\label{sec:Improvement}
Here we examine several methods of reducing the contrast defect.

\subsection{Better polishing}
The most direct approach is to reduce the mirror figure
error. However, reaching appropriate levels of surface roughness is
beyond the capabilities of current technology. We estimate that in
order to achieve reasonable performance, LG$_{3,3}$ modes require
mirrors with an RMS roughness roughly one order of magnitude smaller
than is currently achievable (assuming the mirror coatings introduce
no additional roughness, i.e.\ perfectly smooth, uniform coatings).

\begin{figure}
    \includegraphics[width=\columnwidth]{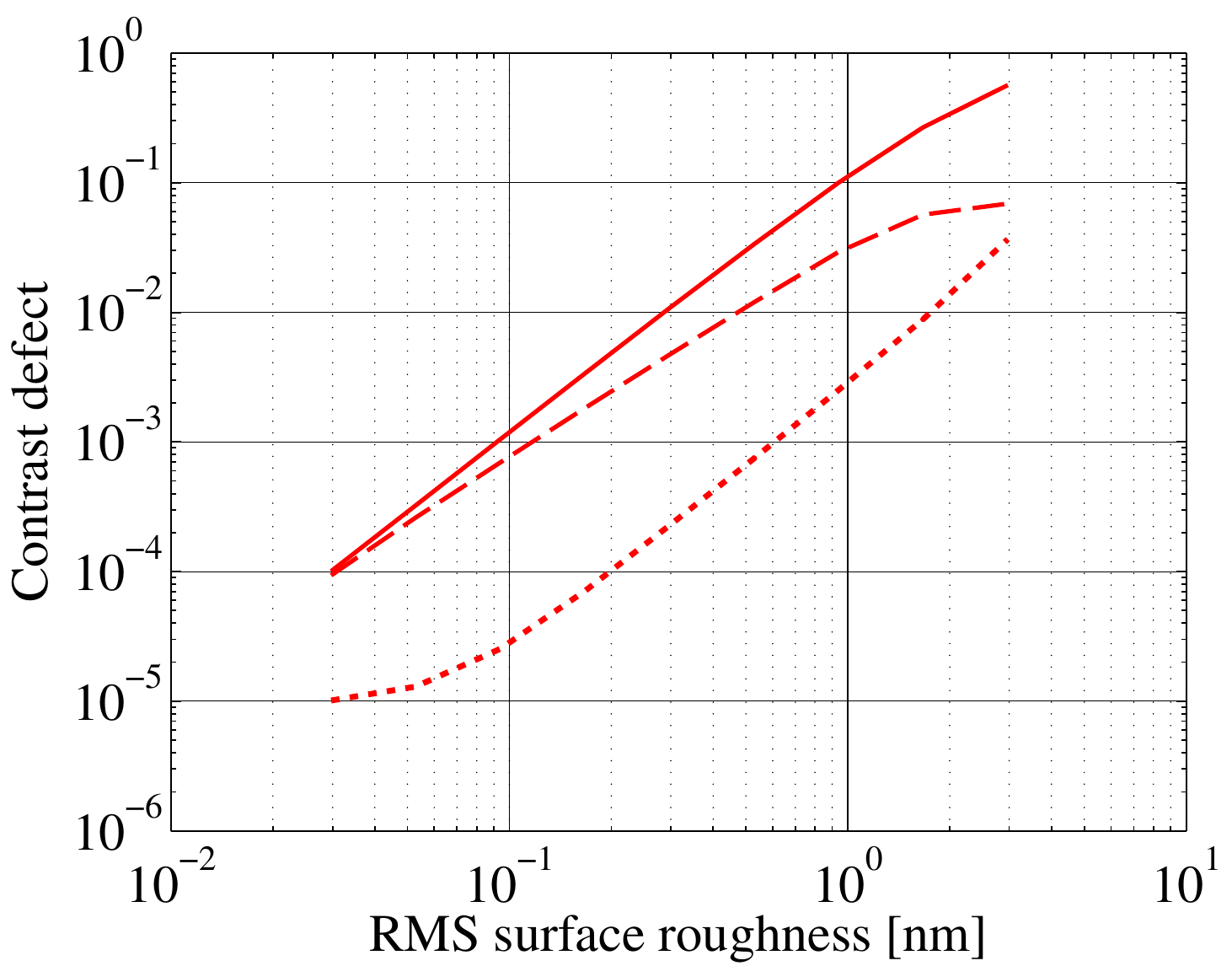}
    \caption{Analytic calculation with different
      conditions for reducing the contrast defect: solid curve is the original contrast defect; dashed line has
      corrective rings added to the phase map; dotted curve is with detuned injecting laser frequency.}
    \label{fig:analyticalplot}
\end{figure}

In the remainder of this section, we thus consider more unconventional
means of reducing the contrast defect. Fig.~\ref{fig:analyticalplot}
shows the results of each investigation.

\subsection{Arm cavity detuning}
Equation~(\ref{eq:inoutrelation}) shows that the output field varies
with the frequency of light injected into a cavity, or equivalently
with cavity length. This motivated us to study the variation in
contrast defect as a function of arm cavity detuning.

It was found that detuning was effective in modifying contrast defect.
Unfortunately, the large detuning necessary to recover good contrast
had the effect of simultaneously reducing the optical power
circulating in the cavity.

With a detuning of 100~Hz (approximately two cavity linewidths), it was
possible to recover acceptable contrast (dotted line,
Fig.~\ref{fig:analyticalplot}). However, the same detuning causes the
circulating power to drop by $\sim$60\%. We therefore do not consider
this approach to be viable in gravitational-wave interferometers where
the injected frequency is usually tuned to maximize the optical power
circulating in the coupled-cavity system. This technique can, however,
be considered for experiments where the thermal noise needs to be
reduced, but the cavity's stored power is not of concern.

\subsection{Mirror Corrections}
\label{subsec:Correction}
The increase in contrast defect observed when using LG$_{3,3}$ modes
in the presence of realistic surface roughness results from the
presence of multiple pseudo-degenerate higher order modes. Here we
attempt to see if this effect can be mitigated by depositing
corrective structures on the mirror's surface.

By introducing material at the nodes of the desired LG$_{3,3}$ mode it
was hoped that the unwanted modes from the same sub-space could be
suppressed.

\begin{figure}[htbp!]
    \includegraphics[width=\columnwidth]{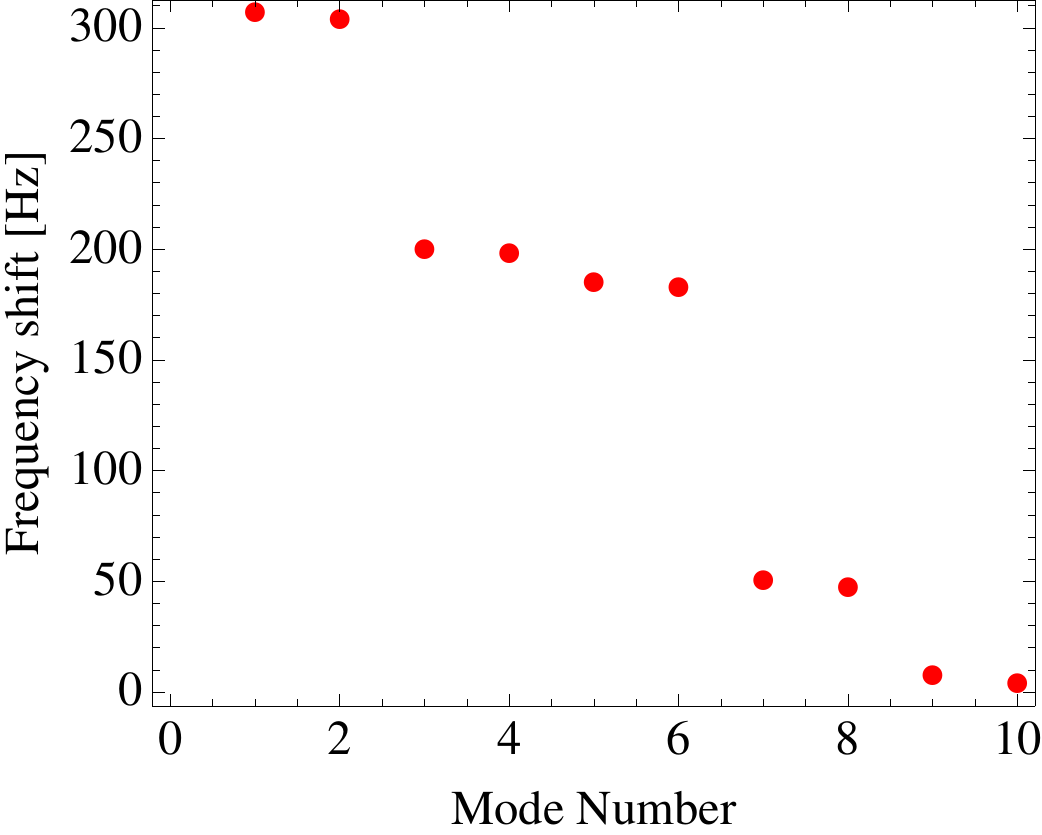}
    \caption{Frequency shift in LG modes introduced by adding rings to
      the reference phase map.}
  \label{fig:ringsplit}
\end{figure}

As a concrete example, two Gaussian rings were added to the random
phase map at nodes 1 and 3 of the LG$_{3,3}$ mode. Each ring was of
the form: \be f(x,y)=\frac{\lambda}{20}
e^{-\frac{(r-r_p)^2}{2(R/100)^2}}.\ee where $r_p$ indicates the
position of the different nodes.  The frequency split is plotted in
Fig.~\ref{fig:ringsplit}. Compared to Fig.~\ref{fig:split}, the
frequency splits under this condition are much larger, therefore the
other degenerate modes will be harder to excite.

The analytically computed contrast defect for this case is plotted as
the dashed line in Fig.~\ref{fig:analyticalplot}. Although the defect
is improved for values of surface roughness similar to that which is
currently achievable, we find that, in order to significantly break
the modal degeneracy, the height of the rings must be increased to
such a degree that the induced scatter becomes unacceptably high
($\sim$~500~ppm), reducing the stored power and thus the
interferometer's phase sensitivity. We hence conclude that this
approach is not promising.

At Caltech, Yamamoto has studied a similar approach whereby the mirror
reflectivity is set to zero at the nodes of the LG$_{3,3}$ mode
\cite{HiroNote}. This technique was found to be similarly unsuitable
for application to gravitational wave interferometers.

\subsection{Mode Healing}
Previous work~\cite{Brett:DR} has shown that the presence of a signal
recycling cavity can substantially reduce the contrast defect in the
case where the resonant mode in the interferometer is TEM$_{0,0}$. The
higher order transverse modes are not resonant in the signal recycling
cavity and are therefore suppressed. In the LG$_{3,3}$ case, however,
the signal cavity is resonant for the LG$_{3,3}$ mode as well as all
of the modes which are in the degenerate sub-space. Therefore, our
expectation is that there {\it would not} be a mode healing effect
when using any higher order mode which can be split in this way.  In
the case where the signal recycling cavity is detuned to amplify the
gravitational wave response at a particular frequency, the situation
could be significantly more complicated due to the frequency splitting
shown in Fig.~\ref{fig:split}. To quantitatively explore the effect of
the compound cavity on degenerate modes, further analytic and
numerical work is required.

\section{Conclusions}
In this paper, we use numerical analysis as well as perturbation
theory to analyze the modes of a Fabry-Perot cavity resonating a
LG$_{3,3}$ beam. We prove that with realistic mirror figure errors,
the real output mode of the cavity will change significantly,
resulting in an unacceptable increase of the contrast defect.

We also investigate unconventional corrective techniques to reduce the
contrast defect.  While they turn out to be unsuitable for quantum
shot noise limited interferometers, they may have some utility for
other classes of cavities.  For LG$_{3,3}$ modes to function
effectively, we estimate that surface figure errors must be reduced to
the order of $10^{-2}$~nm~RMS to achieve the required contrast defect of LIGO($\sim 10^{-4}$). Such precise polishing and coating uniformity will
likely not be available for several years. Using high-order
Laguerre-Gauss modes in standard spherical mirror cavities appears to
be a poor choice in light of current technologies.

Numerical simulations using mesa and normal Gaussian beams show 
these beams are not so sensitive to figure errors.
Future effort will be directed toward the construction of a new family
of optical modes which can reduce the thermal noise impact while
simultaneously being robust against mirror imperfections.

\section{Acknowledgements}
We thank Andreas Freise, Matteo Barsuglia, and Bill Kells for several illuminating
discussions. Many thanks to Eric Gustafson and Huan Yang for a careful
reading of this manuscript. TH and YC are supported by 
NSF Grant PHY-0601459, PHY-0653653, CAREER Grant PHY-0956189, and
the David and Barbara Groce Startup Fund at the California Institute of Technology. 
RA and HY are supported by the National Science Foundation under grant PHY-0555406.
JM is the recipient of an Australian Research Council Post Doctoral Fellowship
(DP110103472).

\appendix

\section{Contrast Defect}
\label{sec:Equiv}
Here we show that the contrast defects defined in
equations \eqref{eq:ContrastDefect} and \eqref{eq:NumCD} are
equivalent in the limit of small perturbations.  We denote the input
and output field for the two cavities (X and Y) as $\ket{{\rm in}}_X,
\ket{{\rm out}}_X, \ket{{\rm in}}_Y, \ket{{\rm out}}_Y$.
In our analytical calculation, we assume the input and output fields of each cavity are normalized, because we ignore
transmissivity of the ETM, and the diffraction loss, so that we have
\be
\braket{{\rm in}}{{\rm in}}_X=\braket{{\rm in}}{{\rm in}}_Y=\braket{{\rm out}}{{\rm out}}_X=\braket{{\rm out}}{{\rm out}}_Y=1.
\ee
We can then write 
\be
P_X=\braket{{\rm in}}{{\rm in}}_X=1, P_Y=\braket{{\rm in}}{{\rm in}}_Y=1
\ee

and the power at the anti-symmetric port can be written as
\ba
P_{AS}&=&\|\ket{{\rm out}}_X-\ket{{\rm out}}_Y\|^2\nonumber\\
&=&2-\braket{{\rm out}}{{\rm out}}_{XY}-\braket{{\rm out}}{{\rm out}}_{YX}.
\ea
With the definition of contrast defect in Eq.~(\ref{eq:ContrastDefect}), we have
\be
\epsilon_X=1-|\braket{{\rm in}}{{\rm out}}_X|, \epsilon_Y=1-|\braket{{\rm in}}{{\rm out}}_Y|
\ee
so that we can write the output field as
\ba
\ket{{\rm out}}_X= (1-\epsilon_X)e^{i\phi_X}\ket{\rm in}_X+\ket{\delta_X}\nonumber \\
\ket{{\rm out}}_Y= (1-\epsilon_Y)e^{i\phi_Y}\ket{\rm in}_Y+\ket{\delta_Y}
\ea
where
\ba
\braket{\delta_X}{{\rm in}}_X=0, \braket{\delta_X}{\delta_X}=\epsilon_X\nonumber\\
\braket{\delta_Y}{{\rm in}}_Y=0, \braket{\delta_Y}{\delta_Y}=\epsilon_Y\nonumber\\
\braket{\delta_X}{{\rm in}}_Y=\braket{\delta_Y}{{\rm in}}_X=0
\ea

In both the analytical and numerical calculations, we assume one of the two interferometer cavities is perfect and the other
is with mirror figure errors, so here we can write $\epsilon_Y=1, \ket{\delta_Y}=0$, therefore, to the first order approximation ($\epsilon_X\ll1$), we get obtain
\be
 P_{AS}=2\epsilon_X
\ee
which shows that the contrast defect defined in Eq.~\eqref{eq:NumCD} is the same as the analytical definition in Eq.~\eqref{eq:ContrastDefect}.

When we consider two cavities both with imperfections, if they are statistically independent, we can write
$\braket{\delta_X}{\delta_Y} = 0$, so that the average value of the contrast defect of the system is
\be
 P_{AS}=2(\epsilon_X+\epsilon_Y).
\ee

We now show that Eq.~(\ref{eq:ContrastDefect}) can be approximately
written as given in Eq.~(\ref{eq:ContrastDefectApprox}). From
Eq.~(\ref{eq:output}), we have
\be
\label{ovlap}
\braket{{\rm in}}{{\rm out}}=\sum_{n'} \frac{\gamma+i (\omega-{\omega_n'})}{\gamma-i (\omega-\omega_{n'})}\braket{{\rm in}}{n'} \braket{n'}{{\rm in}}
\ee
when $\omega-\omega_{n'}\ll \gamma$, Eq.~\eqref{ovlap} can be expanded as
\be
\sum_{n'} \left[1+\frac{2i (\omega-\omega_{n'})}{\gamma}-\frac{2(\omega-\omega_{n'})^2}{\gamma^2}\right]\braket{{\rm in}}{n'} \braket{n'}{{\rm in}}
\ee

Note that when the mode frequency is shifted $\omega_{n'}$, the optical power in the cavity is maximized, so that the
linear term vanishes, therefore that the modulated frequency of the beam
$\omega$ can be given by:
\be
\label{omega}
\omega=\frac{k c}{L}\bra{33}\delta h\ket{33}
\ee

The contrast defect defined in Eq.~\eqref{eq:ContrastDefect} is
\ba
 \epsilon&=&\sum_{n'}\frac{2(\omega-\omega_{n'})^2}{\gamma^2}\braket{{\rm in}}{n'} \braket{n'}{{\rm in}}\nonumber\\
 &=&\frac{2}{\gamma^2}\left(\sum_{n'} \braket{{\rm in}}{n'}\omega_{n'}^2 \braket{n'}{{\rm in}}-\omega^2\right)
\ea
After perturbation, the frequency split is the eigenvalue
of the matrix given in Eq.~\eqref{eq:matrix} and the eigenvector of
the matrix is the real eigenmode of the cavity, thus we can write
\ba
 &&\sum_{n'} \braket{{\rm in}}{n'}\omega_{n'}^2 \braket{n'}{{\rm in}}\nonumber\\
 &&=\left(\frac{2\pi c}{\lambda {\rm L}}\right)^2\sum_{n'm'l'}\braket{{\rm in}}{n'}\bra{n'}\delta h\ket{m'}\bra{m'}\delta h\ket{l'}\braket{l'}{{\rm in}}\nonumber\\
 &&=\left(\frac{2\pi c}{\lambda {\rm L}}\right)^2\sum_{m'}\bra{{\rm in}}\delta h\ket{m'}\bra{m'}\delta h\ket{{\rm in}}
\ea
where $\gamma=c{T}/4{\rm L}$. Then, in this limit, when the injected field is
$\ket{33}$, the contrast defect can be written as in Eq.~\eqref{eq:ContrastDefectApprox}.


\begin{thebibliography}{99}
\bibitem{Bondu} This algorithm is based on techniques developed by
  F.~Bondu (Observatoire de la C\^{o}te d'Azur/ Institut de Physique de Rennes).
\end{thebibliography}
\end{document}